\documentclass[journal]{IEEEtran}

\usepackage{graphicx}
\usepackage{multirow}
\usepackage{amsmath, amsthm, amssymb}
\usepackage{algorithmic}
\usepackage{algorithm}
\usepackage{float}
\usepackage{subfig}
\usepackage{tikz}
\usepackage{setspace}
\usepackage[switch]{lineno}
\usepackage{multicol}
\usepackage{mathtools}
\usepackage{hyperref}

\begin{document}

\title{Learning deep pyramid based representations for pansharpening}

\author{Hannan Adeel, Syed Sohaib Ali, Muhammad Mohsin Riaz, Syed Abdul Mannan Kirmani, Muhammad Imran Qureshi and Junaid Imtiaz 
\IEEEcompsocitemizethanks{\IEEEcompsocthanksitem Hannan Adeel is currently pursuing his PhD from Electrical Engineering department, COMSATS University Islamabad, Pakistan.  \protect
E-mail: dr.hannanadeel@gmail.com
\IEEEcompsocthanksitem  Syed Sohaib Ali is serving as Assistant Professor in Computer Science department, COMSATS University Islamabad, Pakistan.  \protect
E-mail: sohaib.ali@comsats.edu.pk
\IEEEcompsocthanksitem  Muhammad Mohsin Riaz is serving as Assistant Professor in Center for Advanced Studies in Telecommunication (CAST), COMSATS University Islamabad, Pakistan.
\IEEEcompsocthanksitem  Syed Abdul Mannan Kirmani is serving as Assistant Professor in Center for Advanced Studies in Telecommunication (CAST), COMSATS University Islamabad, Pakistan.
\IEEEcompsocthanksitem  Muhammad Imran Qureshi is serving as Assistant Professor in Dept of Electrical Engineering, FICT, BUITEMS, Quetta, Pakistan.
\IEEEcompsocthanksitem  Junaid Imtiaz is serving as Head, Dept of Electrical Engineering, Bahria University Islamabad, Pakistan.
}
\thanks{}}

\markboth{}%
{Shell \MakeLowercase{\textit{Hannan  et al.}}: Learning deep multiresolution representations for pansharpening}

\IEEEtitleabstractindextext{
\begin{abstract}
Retaining spatial characteristics of panchromatic image and spectral information of multispectral bands is a critical issue in pansharpening. This paper proposes a pyramid based deep fusion framework that preserves spectral and spatial characteristics at different scales. The spectral information is preserved by passing the corresponding low resolution multispectral image as residual component of the network at each scale. The spatial information is preserved by training the network at each scale with the high frequencies of panchromatic image alongside the corresponding low resolution multispectral image. The parameters of different networks are shared across the pyramid in order to add spatial details consistently across scales. The parameters are also shared across fusion layers within a network at a specific scale. Experiments suggest that the proposed architecture outperforms state of the art pansharpening models. The proposed model, code and dataset is publically available at \href{https://github.com/sohaibali01/deep_pyramid_fusion}{github} \footnote{\url{https://github.com/sohaibali01/deep_pyramid_fusion}}.

\end{abstract}

 }

\maketitle

\IEEEdisplaynotcompsoctitleabstractindextext

\IEEEpeerreviewmaketitle

\section{Introduction}\label{sec:introduction}

\begin{figure}[h]
\centering
{\includegraphics[scale=0.8]{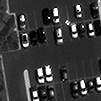}}
{\includegraphics[scale=0.8]{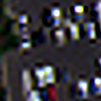}}
{\includegraphics[scale=0.8]{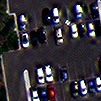}}
\caption{\footnotesize Pansharpening example on WorldView-3 Satellite Imagery: Left: Panchromatic Image acquired at $0.31m$ spatial resolution, Middle: Interpolated multispectral image (originally acquired at  $1.24m$ spatial resolution), Right: Fused Output }
\label{fig:pansharpening}
\end{figure}

\IEEEPARstart{S}{atellite} imagery is mostly acquired from two types of sensors for the purpose of remote sensing. The panchromatic sensor provides a single channel high resolution panchromatic image (HR-PAN) that has high spatial resolution but lacks spectral colors. The second sensor provides a low resolution multispectral image (LR-MSI) containing several bands but lacks high spatial resolution. Combining spatial and spectral characteristics of both sensors (as shown in figure \ref{fig:pansharpening}) is a challenging task in pansharpening. The process of pansharpening aims to produce a high-resolution multispectral image (HR-MSI) by fusing high spatial resolution of panchromatic  image and rich spectral information of multispectral image. This results in a tradeoff between spectral and spatial information in the output HR-MSI \cite{loncan2015hyperspectral}.

\normalsize


\section{Literature Review}
The traditional pansharpening methods can be classified into three classes; 
spectral transformation based component substitution (CS) methods, spatial transformation based multi-resolution analysis (MRA) methods and model based methods.

In CS based methods, components such as intensity obtained through spectral transformation of LR-MSI is merged with HR-PAN. Intensity hue saturation (IHS) \cite{ghahremani2016nonlinear}, \cite{li2019model}, multivariate statistical methods such as principal component substitution (PCS) \cite{rahimzadeganasl2019approach}, Gram-Schmidt (GS) spectral transformation \cite{Li2020improving}, and band-dependent spatial-detail (BDSD) schemes \cite{vivone2019robust} are representative methods of this class. The non-linear (NL) IHS method is proposed in \cite{ghahremani2016nonlinear} to optimize the intensity component. In \cite{li2019model} Fast generalized IHS (FGIHS) method uses variational optimization to increase the accuracy of estimated intensity image in the IHS space and to minimize the associated spectral distortion. The authors in \cite{rahimzadeganasl2019approach} recently proposed CS based approach to target HR-MSI by replacing the histogram matched luminance component of upscaled LR-MSI represented in $CIELab$ color space with original HR-PAN. In \cite{Li2020improving} spectral sensitivity is used in conjunction with IHS and Gram-Schmidt transformation. These methods achieve comparable spatial sharpness and improved spectral information in the fused image as compared to conventional CS methods. \textit{Vivone} \cite{vivone2019robust} recently proposed regression and physical constraints based variants of band-dependent spatial-detail (BDSD) approach. In general, CS based methods suffer from distortion owing to the difference between estimated component values and PAN data. %

The MRA based methods commonly use multiresolution representations such as Laplacian pyramids (LP) \cite{9024138}, wavelet family transforms \cite{wady2020new}, non-subsampled transforms \cite{lu2017improved}-{\cite{jiao2019pansharpening}, latent low rank (LLR) decomposition \cite{hallabia2019pan} and morphological pyramids (MP) \cite{restaino2016fusion} to extract information from spatially enriched PAN image. MRA based methods are generally superior in terms of spectral preservation but artifacts appear due to subsampling in the PAN image \cite{meng2019review}. Regression based generalised LP framework proposed in \cite{9024138} achieves high quality results on different satellite imagery with heavy computational burden. MRA based method proposed in \cite{wady2020new} decomposes HR-PAN image using non-decimated "$\grave{a}$trous" wavelet transform (ATWT) to produce low resolution PAN image (LR-PAN). Then linearly weighted LR-PAN and re-sampled multispectral image is used to achieve color preservation. In \cite{lu2017improved} non-subsampled contourlet transform (NSCT) has shown comparable performance to BDSD. \textit{Kumar} \cite{kumar2019pan} uses non subsampled shearlet transform (NSST) in order to avoid frequency aliasing and achieving better directional selectivity and shift invariance. In \cite{jiao2019pansharpening}, NSST is used with guided filtering based MRA approach, however the process introduces blurry edges consequently affecting spatial quality. MRA of LLR based composite image is proposed in \cite{hallabia2019pan}. This framework extracts details from reconstructed image using LLR decomposition of HR-PAN and LR-MSI data. In \cite{restaino2016fusion} morphological gradients have been used in a nonlinear pyramidical setting. This scheme shows good results for various satellite imagery. Recently in \cite{li2020image} image segmentation based pansharpening method is proposed to reduce the effect of spectral distortion caused by fused spectra of mixed pixels in MRA based methods.

In addition to CS and MRA based methods, model based methods pose pansharpening as a local optimization problem. In \cite{chen2019pan} sparse representation and low rank regularization is used in a variational optimization setup however this scheme requires empirical parameter setting and suffers from computational cost for high quality pan sharpening. \textit{Khatheri et al.} \cite{khateri2019model} proposed a model based pan sharpening that uses sparse coefficients with patch dictionary to generate HR-MSI. The fused output preserves image and patch energy ratio as that of LR-MSI. In \cite{wang2019regularised} spatial information regularization is carried out to combat local dissimilarities by optimizing a convex energy function. Spectral consistency based variational optimization model based on half quadratic optimization \cite{yang2009fast} presented in \cite{khateri2019regularised} aims to reduce spectral distortion during pansharpening. This model shows good spectral preservation while also preserving spatial smoothness of HR-PAN. \textit{Tian et al.} \cite{tian2020variational} reported that a general assumption regarding sparse representation based methods is that multi-resolution images are represented by same sparse coding under some dictionaries. They proposed gradient sparse coding based variational model and used gradient similarity based pansharpening. In \cite{wang2018high}, Bayesian model is used to jointly express the LR-MSI, HR-PAN and HR-MSI using multiorder gradients (MoG) optimized by alternating direction method of multipliers (ADMM). In \cite{yang2018pansharpening} spectral and intensity modulation coefficients obtained from spectral and statistical measures of LR-MSI and HR-PAN are combined in an adaptive linear model to strike a balance between spectral and spatial details in target HR-MSI. \textit{Fu et al.} \cite{fu2019variational} formulated HR-MSI  problem as a variational optimization problem based on regression and gradient difference of HR-PAN and LR-MSI for spatial preservation. In general, model based methods focus on certain aspects and have high computational cost due to local optimization solvers.

Deep learning (DL) based methods assume non-linear mapping between LR-MSI and HR-PAN images. Such methods can be used both in supervised and unsupervised setting. Recently \textit{Guo et al.} \cite{guo2020bayesian} proposed mulitscale recursive block based convolutional neural network (CNN) trained in MOG domain for pansharpening. Residual blocks with multi-scale LP and parameter sharing along network branches is used in \cite{jiang2020learning} to improve fusion performance. In \cite{8668404}, convolutional auto-encoders (CAE) learns the mapping between LR and HR space using degraded PAN and HR-PAN patches. Afterwards CAE based estimated HR-MSI is used to preserve spectral-spatial details in the target HR-MSI. In \cite{li2019cnn} the mapping is estimated by CNN based on pyramid structure to minimize the spectral and spatial dissimilarity in HR-PAN and LR-MSI. \textit{Masi et al.} \cite{masi2016pansharpening} proposed CNN based pansharpening (PNN) method using three-layered architecture. Their stacked network uses interpolated LR-MSI along with HR-PAN image allowing training at the target resolution. \textit{Scarpa et al.} \cite{scarpa2018target} proposed the target-adaptive enhanced version of PNN acronym as PNNplus or PNN+. The deeper network is trained to produce residual pansharpened image corrected by up-sampled LR-MSI through skip connections. In comparison to PNN \cite{masi2016pansharpening} this network achieves better performance using $L1$ loss function for different types of satellite imagery. Inspired by residual network (ResNet)\cite{he2016deep}, \textit{Wei et al.} \cite{wei2017boosting} proposed deep residual PNN (DRPNN) with multiple sparse residual layers to boost the network efficiency. Pansharpening network (PanNet) \cite{yang2017pannet} is a ResNet based DL scheme that incorporates up-sampled LR-MSI for spectral correction in the target super resolved image and high frequency components for edge preservation and avoiding inconsistencies between HR-PAN and HR-MSI image data. PanNet shows good quality fusion for variety of satellite imagery. \textit{Liu et al.} \cite{liu2018psgan} and  \textit{Ma et al.} \cite{ma2020pan} demonstrated the use of generative adversarial networks (GANs) for pansharpening. \textit{Lou et al.} \cite{luo2020pansharpening} formulated a new loss function based on original HR-PAN and LR-MSI input pair and target HR-MSI, achieving good spatio-spectral performance for small scale objects. Differential Information residual CNN \cite{jiang2020differential} learns mapping between HR-PAN and LR-MSI and HR-PAN and HR-MSI based on residual block and attention modules to fully cascade high and low frequency components and feature refinement. Deep leaning with CS method is proposed in \cite{qu2020unsupervised} where stacked self-attention modules are used and sub-pixel level spectral details are injected in LR-MSI to obtain target HR-MSI. Similarly \textit{Ozcelik et al.} \cite{ozcelik2020rethinking} have shown in their PSColorGAN that along with color injection, adopting random scale down-sampling strategy during training enhances the overall performance of the network and minimizes the blurring effect. 

\section{Proposed Scheme}
Since most of the satellite sensors including Worldview-4, Ikonos, QuickBird etc provide multispectral image that is acquired at a quarter of a spatial resolution compared to the panchromatic band, the proposed pyramid framework in figure \ref{fig:proposed} uses a two stage decomposition and reconstruction. The proposed pansharpening procedure inherently assumes that,
\begin{itemize}
  \item The low frequencies of the output fused image can be directly obtained by the input multispectral image.
  \item At each scale, the high frequencies of the output can be estimated by the corresponding high frequencies of panchromatic band but this estimation should be coherent with the low frequencies obtained at the coarser scale.
  \item The spatial consistency assumption should be followed. For instance, once the output multispectral image is downsampled from $1m$ to $2m$, the residual high frequencies should be proportional to the ones obtained if the current image is again downsampled from $2m$ to $4m$. Figure \ref{fig:proposed} ensures this by sharing the network parameters across two levels of the pyramid.
\end{itemize}

\begin{figure*}[t]
\centering
{\includegraphics[scale=0.9]{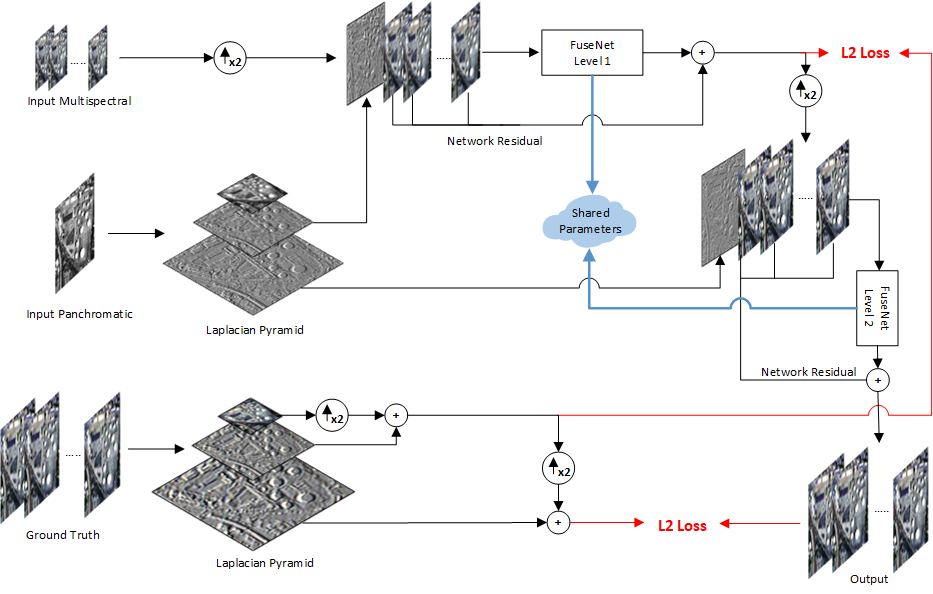}}
\caption{Proposed architecture for pansharpening}
\label{fig:proposed}
\end{figure*}

\begin{figure*}[t]
\centering
{\includegraphics[scale=0.8]{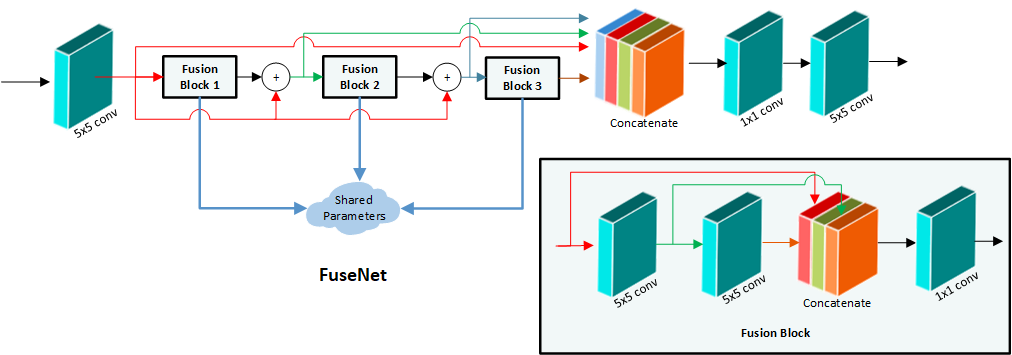}}
\caption{Proposed Network}
\label{fig:network}
\end{figure*}

\begin{figure*}[t]
\centering
\subfloat[\footnotesize Original Image]{\includegraphics[scale=1]{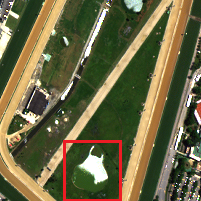}} \hspace{2pt}
\subfloat[\footnotesize PNN \cite{masi2016pansharpening}]{\includegraphics[scale=1]{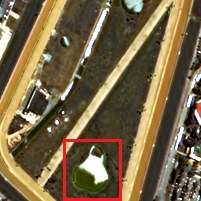}}  \hspace{2pt}
\subfloat[\footnotesize PNN+ \cite{scarpa2018target}]{\includegraphics[scale=1]{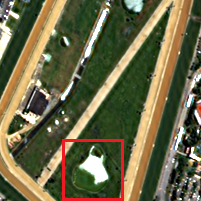}} \\
\subfloat[\footnotesize DRPNN \cite{wei2017boosting}]{\includegraphics[scale=1]{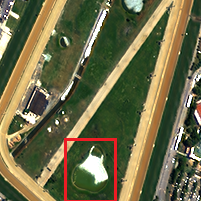}}  \hspace{2pt}
\subfloat[\footnotesize PanNet \cite{yang2017pannet} ]{\includegraphics[scale=1]{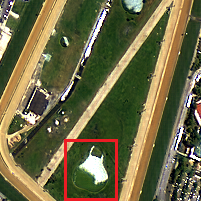}}  \hspace{2pt}
\subfloat[\footnotesize VPLG \cite{fu2019variational} ]{\includegraphics[scale=1]{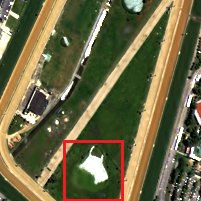}}  \hspace{2pt}
\subfloat[\footnotesize Proposed FuseNet]{\includegraphics[scale=1]{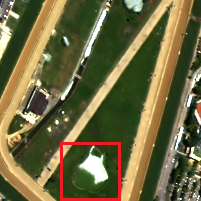}}
\caption{\footnotesize Worldview 2 Pansharpening}
\label{fig4:results}
\end{figure*}
\normalsize

\begin{figure*}[t]
\centering
\subfloat[\footnotesize PNN difference]{\includegraphics[scale=0.4]{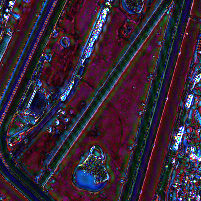}}  \hspace{1pt}
\subfloat[\footnotesize PNN+ difference]{\includegraphics[scale=0.4]{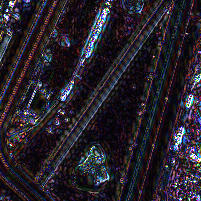}} \hspace{1pt}
\subfloat[\footnotesize DRPNN difference]{\includegraphics[scale=0.4]{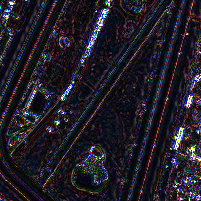}}  \hspace{1pt}
\subfloat[\footnotesize PanNET difference ]{\includegraphics[scale=0.4]{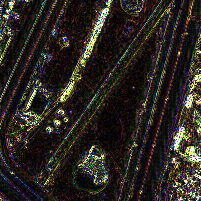}}  \hspace{1pt}
\subfloat[\footnotesize VPLG difference ]{\includegraphics[scale=0.4]{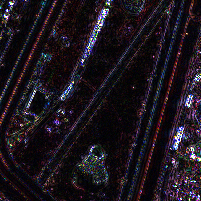}}  \hspace{1pt}
\subfloat[\footnotesize Proposed difference]{\includegraphics[scale=0.4]{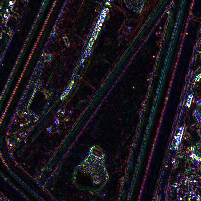}}
\caption{\footnotesize Absolute Difference between input and different output images presented in figure \ref{fig4:results}}
\label{fig5:diffResults}
\end{figure*}
\normalsize

\subsection{Pyramid based fusion}
Let the input panchromatic image $\left(P\right)$ be decomposed into $J$ levels using laplacian pyramids such that,
\begin{equation}\label{eq:1}
\overline{P}_{j}=\Big(\overline{P}_{j-1}\ast h\Big) \downarrow_2 \hspace{10pt} \forall \hspace{5pt} j\in\{1,2,...,J\}
\end{equation}
where $\overline{P}_{j}$ represents a low pass output of $j^{th}$ level. For initialization, $\overline{P}_0=P$, while $h$ can be a symmetric and separable low pass filter such as that used in \cite{burt1983laplacian}. The high frequencies at $j^{th}$ level are estimated as an input-output difference, i.e.,
\begin{equation}\label{eq:2}
\hat{P}_{j}=\overline{P}_{j-1} - \Big(\overline{P}_{j}\uparrow_2\Big)
\end{equation}
Similarly let the $j^{th}$ approximation of the $b^{th}$ band of the multispectral image ($\overline{M}_{b}$) be represented as $\overline{M}_{b,j}$. The upsampled version of the input multispectral image provides the approximation for the $1^{st}$ scale, i.e.,
\begin{equation}\label{eq:3}
\overline{M}_{b,1}=\Big(\overline{M}_{b}\uparrow_2\Big)
\end{equation}
At each scale the corresponding high frequencies of the panchromatic image are stacked together with low pass approximations of the coarser scale and the image stack is passed through the fusion network to obtain the corresponding scale approximation of the fused image, i.e.,
\begin{equation}\label{eq:4}
S_j = \Big[ \hat{P}_{j}, \overline{M}_{1,j}, \overline{M}_{2,j}, ... , \overline{M}_{B,j} \Big]
\end{equation}
where $B$ is the total number of input multispectral bands. $S_j$ represents image stack at $j^{th}$ scale obtained by channelwise concatenation of high frequency and low frequency bands. This stack is then passed through the fusion network that estimates the $j^{th}$ scale approximation of the fused image.
\begin{equation}\label{eq:recursive}
\overline{M}_{b,j+1} = \Big( f_{\text{FuseNet}}(S_j) + \overline{M}_{b,j} \Big) \uparrow_2
\end{equation}
where $f_{\text{FuseNet}}(\cdot)$ provides a deep convolutional mapping discussed in the next section. Note that this mapping doesn't vary across scales due to parameter sharing. The skip connection in the form of $\overline{M}_{b,j}$  in eq. (\ref{eq:recursive}) preserves the identity mapping and allows the framework to output the interpolated version of the low resolution input multispectral image in the worst case scenario. The output image $\overline{M}_{b,J}$ is obtained via eq. (\ref{eq:recursive}) in recursive fashion.

Given a ground truth image, $M_{b}$, the parameters of $f_{\text{FuseNet}}$ are optimized by minimizing the following loss function,
\begin{equation}\label{eq:loss}
\mathcal{L}_{2} = \sum_j {\| M_{b,j} - \overline{M}_{b,j} \|} ^ 2
\end{equation}
where $M_{b,j}$ represents $j^{th}$ scale approximation of ground truth obtained by summing low and high frequency components of the pyramid,
\begin{equation}\label{eq:7}
M_{b,j} = \hat{M}_{b,j} + \overline{M}_{b,j}
\end{equation}
where $\overline{M}_{b,j}$ and $\hat{M}_{b,j}$  represent low and high pass components respectively which are obtained by following eq. (\ref{eq:1}) and eq. (\ref{eq:2}) respectively.

\begin{figure*}[t]
\centering
\subfloat[\footnotesize Panchromatic Image]{\includegraphics[scale=0.6]{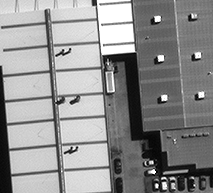}}
\subfloat[\footnotesize Interpolated Multispectral Image]{\includegraphics[scale=0.6]{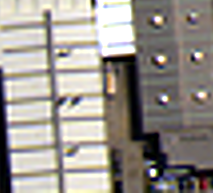}}
\subfloat[\footnotesize PNN \cite{masi2016pansharpening}]{\includegraphics[scale=0.6]{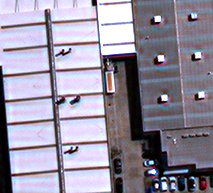}}
\subfloat[\footnotesize PNN+ \cite{scarpa2018target}]{\includegraphics[scale=0.6]{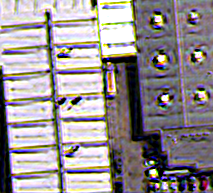}} \\
\subfloat[\footnotesize DRPNN \cite{wei2017boosting}]{\includegraphics[scale=0.6]{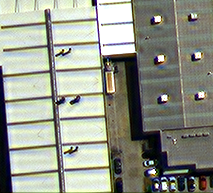}}
\subfloat[\footnotesize PanNET \cite{yang2017pannet} ]{\includegraphics[scale=0.6]{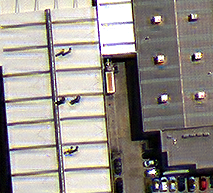}}
\subfloat[\footnotesize VPLG \cite{fu2019variational} ]{\includegraphics[scale=0.6]{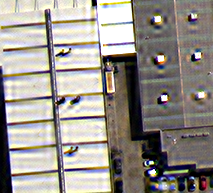}}
\subfloat[\footnotesize Proposed FuseNet]{\includegraphics[scale=0.6]{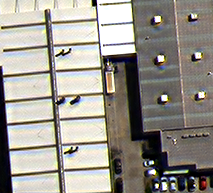}}
\caption{\footnotesize Full scale Worldview 3 Pansharpening}
\label{fig6:results}
\end{figure*}
\normalsize

\subsection{Proposed FuseNet}
The proposed fully convolutional FuseNet (shown in figure \ref{fig:network}) provides a residual learning framework while simultaneously combining local and global features in a hierarchical fashion.

The output $F_n$ at $n^{th}$ layer is calculated as,
\begin{equation}\label{eq:8}
F_n = \sigma \Big( W_n \ast F_{n-1} + b_n \Big)
\end{equation}
where $W_n$ and $b_n$ represent filter weights and bias at $n^{th}$ layer respectively while $\ast$ and $\sigma$ represents convolution and activation function respectively. The first layer extracts a shallow representation from $j^{th}$ input stack $S_j$, i.e., $F_0 = S_j$. The output $F_1$ is then passed onto the stack of $K$ residual learning blocks, each of which learns the local features for fusion.

\subsection{Local Feature Fusion Block}
The local feature fusion is obtained by passing the input of the block to two consecutive $5\times5$ convolutional layers and then concatenating the two layers alongside the input. The final output feature map of the block is estimated by linear weighting of the concatenated channels which is done using $1\times1$ convolution.

Let the output of $n^{th}$ layer at $k^{th}$ local fusion block be represented as  $F_{k,n}$, such that,
\begin{equation}\label{eq:9}
F_{k,n} = \sigma \Big( W_{k,n} \ast F_{k,n-1} + b_{k,n} \Big)
\end{equation}
Since the filter weights are biases are shared across $K$ blocks, we have,
\begin{equation}\label{eq:10}
 W_{k,n} = W_{k+1,n} = W_{k+2,n} = . . . = W_{K,n}
\end{equation}
Likewise,
\begin{equation}\label{eq:11}
 b_{k,n} = b_{k+1,n} = b_{k+2,n} = . . . = b_{K,n}
\end{equation}
The third layer is channel-wise concatenation of the preceding layers, i.e.,
\begin{equation}\label{eq:12}
F_{k,3} = \big[F_{k,0}, F_{k,1}, F_{k,2} \big]
\end{equation}
The output of $k^{th}$ block \big($F_{k,4}$\big) is finally determined as channel-wise linear weighting of the concatenated layers.

\subsection{Global Feature Fusion}
Just like feature maps fusion within a block, the output feature maps of multiple blocks are also fused in a hierarchical manner. The shallow feature map ($F_1$) is used as a skip connection and passed as an additional input to each local fusion block, i.e.,
\begin{equation}\label{eq:13}
F_{k,0} = F_{k-1,0} + F_1
\end{equation}
Finally the feature output maps of each block are concatenated,
\begin{equation}\label{eq:14}
F = \big[F_{k,4}, F_{k+1,4}, F_{k+2,4}, . . . , F_{K,4} \big]
\end{equation}
The feature maps in $F$ are then linearly weighted via $1\times1$ convolution and passed through another $5\times5$ convolutional layer, which outputs $B$ number of output channels.

\begin{center}
\begin{table*}[t]
\centering
\renewcommand{\arraystretch}{1.5}
\caption{Reduced scale pansharpening quality assessment. Best value according to each metric is highlighted.}
\begin{tabular}{|p{2cm}|p{1.35cm}|p{1.35cm}|p{1.35cm}|p{1.35cm}||p{1.35cm}|p{1.35cm}|p{1.35cm}|p{1.35cm}|}  \hline
  Scheme             & \multicolumn{4}{c||}{WorldView 2 (Average on 171 Images) }  & \multicolumn{4}{c|}{WorldView 3 (Average on 100 Images) }  \\ \cline{2-9}
                                   &  QAVE  &  SAM   & ERGAS   & SCC    &  QAVE  &  SAM   & ERGAS   &  SCC    \\ \hline
PNN \cite{masi2016pansharpening}   & 0.7300 & 5.5836 & 4.0532  & 0.8440 & 0.7711 & 8.3305 & 7.3133  & 0.7888   \\ \hline
PNN+  \cite{scarpa2018target}      & 0.7193 & 5.7245 & 4.2630  & 0.8167 & 0.6385 & 9.4245 & 7.7605  & 0.4968   \\ \hline
DRPNN  \cite{wei2017boosting}      & 0.7569 & 5.1577 & 4.0459  & 0.8539 & 0.8269 & 7.0739 & 5.1512  & 0.8331   \\ \hline
PanNet  \cite{yang2017pannet}      & 0.7458 & 4.7030 & 3.9146  & 0.8394 & 0.8607 & 4.8948 & 4.2456  & 0.8674   \\ \hline
VPLG \cite{fu2019variational}      & 0.7479 & 3.9255 & 3.1048  & 0.8771 & 0.9117 & 4.3710 & 3.2945  & 0.8944  \\ \hline
Proposed   & \textbf{0.7905} & \textbf{3.7099} & \textbf{2.8622}  & \textbf{0.9014} &\textbf{0.9145} &\textbf{3.9535} & \textbf{2.8946}  &\textbf{0.9297} \\ \hline
\end{tabular}
\label{table:1}
\end{table*}
\end{center}

\begin{center}
\begin{table*}[t]
\centering
\renewcommand{\arraystretch}{1.5}
\caption{Full scale pansharpening quality assessment. Best value according to each metric is highlighted.}
\begin{tabular}{|p{2cm}|p{1.35cm}|p{1.35cm}|p{1.35cm}||p{1.35cm}|p{1.35cm}|p{1.35cm}|}  \hline
  Scheme             & \multicolumn{3}{c||}{WorldView 2 (Average on 254 Images) }  & \multicolumn{3}{c|}{WorldView 3 (Average on 177 Images) }  \\ \cline{2-7}
                                 & $Q_\lambda$   & $Q_s$  & $QNR$  & $Q_\lambda$ & $Q_s$   &  $QNR$    \\ \hline
PNN \cite{masi2016pansharpening} & 0.1318        & 0.2189 & 0.6908 & 0.0870      & 0.1044  &  0.8221 \\ \hline
PNN+  \cite{scarpa2018target}    & 0.0829        & 0.1624 & 0.7734 & 0.0558      & 0.1098  &  0.8410 \\ \hline
DRPNN  \cite{wei2017boosting}    & 0.1192        & 0.2207 & 0.7043 & 0.0661      & 0.0662  &  0.8742  \\ \hline
PanNet  \cite{yang2017pannet}    & 0.0779        & 0.1719 & 0.7726 & 0.0402      & 0.0576  &  0.9058  \\ \hline
VPLG \cite{fu2019variational}    & \textbf{0.0315}        & 0.1661 & 0.8100 & \textbf{0.0132}      & 0.0462  &  \textbf{0.9412} \\ \hline
Proposed   & 0.0504  & \textbf{0.1237 }&\textbf{0.8377} & 0.0224 & \textbf{0.0418 } & 0.9372 \\ \hline
\end{tabular}
\label{table:2}
\end{table*}
\end{center}

\section{Experiments}
Every convolutional layer of FuseNet uses $48$ number of feature maps except the last $5\times5$ layer which outputs $B$ number of output channels. The parameters at each layer are initialized using Xavier initialization \cite{glorot2010understanding} while the total loss is minimized using ADAM optimizer \cite{kingma2014adam} with a learning rate of $1e^{-3}$. Training is done with a batch size of $20$ and patch size of $192\times192$. Leaky rectified linear unit (Leaky ReLU) with a negative slope of $0.2$ is used as activation function throughout the network. The tensorflow code alongwith training and testing dataset is available at \href{https://github.com/sohaibali01/deep_pyramid_fusion}{github} \footnote{\url{https://github.com/sohaibali01/deep_pyramid_fusion}}.

\subsection{Reduced Scale Quality Assessment}
Since satellite sensors do not provide a ground truth multispectral image at the same high resolution as that of panchromatic image, a common trend is to consider the available low resolution multispectral image as ground truth and simulate the input panchromatic and multispectral images by following Wald's protocol \cite{loncan2015hyperspectral} \cite{wald1997}. Like \cite{yang2017pannet} the input panchromatic and multispectral images are simulated by downsampling original images with a factor of $4$. The output is then compared against the ground truth multispectral image using SAM \cite{yuhas1992discrimination}, relative dimensionless global error in synthesis (ERGAS) \cite{wald2002data}, universal image quality index averaged over the bands (QAVE) \cite{wang2002universal} and spatial correlation coefficient (SCC) \cite{zhou1998wavelet}.

\subsubsection{8-band Pansharpening}
The current work focusses on sensors that provide $8$ multispectral bands alongwith the panchromatic band. These include Worldview2 and Worldview3 which acquire images at different spatial resolutions. Free sample imagery is collected from the internet and a dataset comprising of $431$ images is created, with each image having dimensions of $512 \times 512$. $254$ samples out of $431$ belong to Worldview2 satellite and the remaining $177$ come from Worldview3 satellite. Instead of training the model separately for these two sensors, we train the model jointly on $160$ out of $431$ images selected randomly. The testing is performed on all of the remaining samples.

The proposed scheme is compared with state of the art deep learning models including Pansharpening with convolutional Neural Networks (PNN) \cite{masi2016pansharpening}, PNN+ \cite{scarpa2018target}, Deep Residual Pansharpening Neural Network (DRPNN) \cite{wei2017boosting} and PanNet \cite{yang2017pannet}. The pre-trained models of all of these schemes are already available online. In addition to deep learning models, recently proposed Variational Pansharpening with Local Gradient constraints (VPLG) \cite{fu2019variational} is also included in comparison.

Figure \ref{fig4:results} presents a sample visual comparison between state of the art pansharpening schemes on Worldview2 satellite image while figure \ref{fig5:diffResults} shows the difference of each output scheme with the reference image. PNN produces color distortion in  figure \ref{fig4:results}(b) which is also confirmed since the difference image in figure \ref{fig5:diffResults}(a) seems biased towards red channel. Visually figure \ref{fig4:results}(e) appears the most crisp image but the corresponding difference image in figure \ref{fig5:diffResults}(d) suggests that PanNET produces extra sharp edges that are not present in the reference image. VPLG produces blurring artifacts highlighted in red box in figure \ref{fig4:results}(f).  Compared with the rest, the proposed scheme in \ref{fig5:diffResults}(f) produces least amount of differential edges.

Table \ref{table:1} quantifies state of the art pansharpening schemes using SAM \cite{yuhas1992discrimination}, ERGAS \cite{wald2002data}, QAVE \cite{wang2002universal} and SCC \cite{zhou1998wavelet}. QAVE and ERGAS measure global distortion while SAM and SCC measure spectral and spatial distortion in the output respectively. Table  \ref{table:1} suggests that the proposed scheme outperforms state of art in minimizing both spatial and spectral distortions.

\subsection{Full Scale Quality Assessment}
The models pretrained on reduced scale images are also tested on input images at full scale. In the absence of ground truth, the input low resolution multispectral image is used as spectral reference and the panshromatic image is used as spatial reference. QNR \cite{alparone2008multispectral} is used as an evaluation metric that doesn't need a ground truth reference. QNR itselt measures global distortion but internally measures spectral distortion ($D_\lambda$) and spatial distortion ($D_s$) separately. The values of $D_\lambda$ and $D_s$ should be low, resulting in high value of QNR in the ideal scenario.

Figure \ref{fig6:results} shows visual comparison of different pansharpening schemes on Worldview3 satellite images at full scale. One can see the artifacts around edges of rooftop most noticeably in PNN+ (figure \ref{fig6:results}d) and VPLG (figure \ref{fig6:results}g). Table  \ref{table:2} suggests that VPLG performs quite close to the proposed scheme. A major difference in table \ref{table:2} is that VPLG minimizes spectral distortion ($D_\lambda$) better while the proposed scheme minimizes spatial distortion ($D_s$) better in comparison with the rest.

\section{Conclusion}
This paper proposes a pyramid based deep pansharpening network that minimizes spatial and spectral distortions in a hierarchial fashion. Each level of the pyramid is trained by stacking the low resolution multispectral image obtained at the previous level and the corresponding high frequencies of the panchromatic image at that particular level. The residual component in the form of low resolution multispectral components preserves the spectral mapping between input and output. Experiments suggest that the proposed multiresolution framework outperforms state of the pansharpening models.

\bibliographystyle{IEEEtran}
\bibliography{mybibfile}

\end{document}